\documentclass[aps,prb,twocolumn,showpacs]{revtex4}
\usepackage{txfonts}
\usepackage{graphicx}

\begin{document}

\title{Possible nodeless superconductivity in the noncentrosymmetric superconductor Mg$_{12-\delta}$Ir$_{19}$B$_{16}$}

\author{Gang Mu, Yue Wang, Lei Shan, and Hai-Hu Wen}\email{hhwen@aphy.iphy.ac.cn }

\affiliation{National Laboratory for Superconductivity, Institute of
Physics and Beijing National Laboratory for Condensed Matter
Physics, Chinese Academy of Sciences, P.O. Box 603, Beijing 100080,
People's Republic of China}

\begin{abstract}
We measured the resistivity, diamagnetization, and low-temperature
specific heat of the newly discovered noncentrosymmetric
superconductor Mg$_{12-\delta}$Ir$_{19}$B$_{16}$. The temperature
dependence of specific heat is consistent with the model of an
isotropic s-wave gap with value $\Delta_0 \approx$ 0.94 meV for the
sample $T_c$ = 5.7 K, and the ratio $\Delta_0/k_BT_c\approx 1.91$
indicates a slightly moderate coupling for the superconductivity.
The correlations among the normal state Sommerfeld constant
$\gamma_n$, the slope $-d\mu_0H_{c2}(T)/dT$ near $T_c$, and the
condensation energy $E_c$ are all consistent with the slightly
moderate coupling picture. Based on the data of phonon contribution,
$T_c$ and the McMillan formula, we obtained an electron-phonon
coupling strength $\lambda_{e-ph} \approx$ 0.66, which suggests that
the superconductivity here is induced by the electron-phonon
coupling.

\end{abstract}
\pacs{74.20.Mn,74.20.Rp, 74.25.Bt, 74.70.Dd} \maketitle

\section{Introduction}

The study on superconductivity in noncentrosymmetric materials has
attracted growing efforts in recent
years\cite{GorkovPRL2001,FrigeriPRL2004,EdelsteinJETP1989,LevitovJETP1985,SamokhinPRB2004}.
For most superconductors, the atomic lattice has a centrosymmetry,
therefore the system is inversion symmetric. The orbital part of the
superconducting order parameter has a subgroup which is confined by
the general group of the atomic lattice. Due to the Pauli's
exclusion rule and the parity conservation, the Cooper pair with
orbital even parity should have anti-parallel spin state, namely
spin singlet, while those having orbital odd parity should have
parallel spin state, i.e., spin triplet. If a system lacks the
centrosymmetry, the enhanced spin-orbital coupling may promote the
formation of pairing with high angular momentum, such as the triplet
pairing. Meanwhile the noncentrosymmetric structure allows for the
existence of a mixture of singlet and triplet pairing. Theoretical
features are anticipated in the noncentrosymmetric
system\cite{HayashiPhysicaC2006}. A nodal gap structure has been
observed in Li$_2$Pt$_3$B showing the possibility of triplet
pairing, while due to weaker spin-orbital
coupling\cite{YuanHQPRL,ZhengGQPRL}, the nodal gap has not been
observed in a material Li$_2$Pd$_3$B with similar structure. It is
thus highly desired to investigate the paring symmetry in more
materials with noncentrosymmetric structure.

The newly discovered superconductor Mg$_{10}$Ir$_{19}$B$_{16}$
(hereafter abbreviated as MgIrB ) (Ref. 9) with superconducting
transition temperature $T_c \approx 5$ K is one of the rare
materials which have the noncentrosymmetry. This material has a
space group of $I\texttt{-}43m$ with large and complex unit cells of
about 45 atoms. To some extent it resembles the system
Li$_2$(Pt,Pd)$_3$B since they have alkaline metals (Li, Mg), heavy
transition elements (Pd, Pt, Ir) and the light element boron.
Theoretically it was shown that the major quasiparticle density of
states (DOS) derives from the $d$ orbital of the heavy transition
elements. In this paper we present a detailed investigation and
analysis on the superconducting properties, such as the energy gap,
pairing symmetry, electron-phonon coupling strength, condensation
energy, etc., in MgIrB. Our results suggest that the
superconductivity in this system is of BCS type with an $s$-wave gap
symmetry and a slightly moderate electron-phonon coupling strength.

\section{Sample Preparation and Characterization}

The samples were prepared in two steps starting from pure elements
of Mg (98.5\%), Ir (99.95\%) and, B (99.999\%) using a standard
method of solid state reaction. Appropriate mixtures of these
starting materials were pressed into pellets, wrapped in Ta foil,
and sealed in a quartz tube with an atmosphere of 95\% Ar/ 5\%
H$_2$. The materials were then heated at 600 $^\circ$C and 900
$^\circ$C for 40 min and 80 min, respectively. After cooling down to
room temperature, the samples were reground and then they were
pressed into pellets and sealed in a quartz tube with the same
atmosphere as used in the first step. Some of the time they were
mixed with another certain amount of Mg up to 20\%. In this process
the sample was heated up to 900 $^\circ$C directly and maintained
for 80 min. Usually the superconducting transition becomes sharper
after the second process, but sometimes the $T_c$ may become
slightly lower than the first time. The synthesizing process here is
similar to the previous work reported by the Princeton
group\cite{Klimczuk} but still with some differences. For example,
we used Mg powder instead of flakes to make the mixture more
homogeneous. In addition, the pressure in the sealed quartz tube may
rise to nearly 4 atm at 900 $^\circ$C, which may considerably reduce
the volatilization of Mg during the synthesis. This is probably the
reason that the sample with the transition temperature as high as
$T_c$ = 5.7 K was made without adding extra Mg in the second step
with the starting ratio in the first step as Mg:Ir:B=12:19:16.

The resistivity and the ac susceptibility were measured based on an
Oxford cryogenic system (Maglab-Exa-12). The specific heat was
measured on the Quantum Design instrument physical property
measurement system (PPMS) with temperature down to 1.8 K and the
PPMS based dilution refrigerator (DR) down to 150 mK. The
temperatures of both systems have been well calibrated showing
consistency with an error below 2\% in the temperature range from
1.8 K to 10 K.

\begin{figure}
\includegraphics[width=8cm]{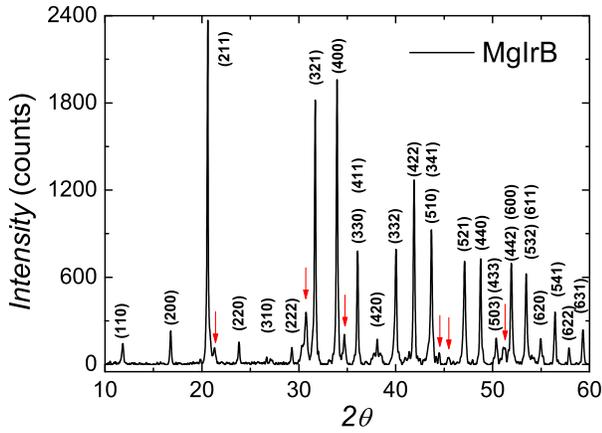}
\caption {(Color online) The x-ray diffraction pattern measured for
the sample $T_c$ = 5.7 K. The peaks from the secondary impurity
phase are marked by the arrows. It is clear that the main
diffraction peaks are from the phase MgIrB.  } \label{fig1}
\end{figure}

\begin{figure}
\includegraphics[width=8cm]{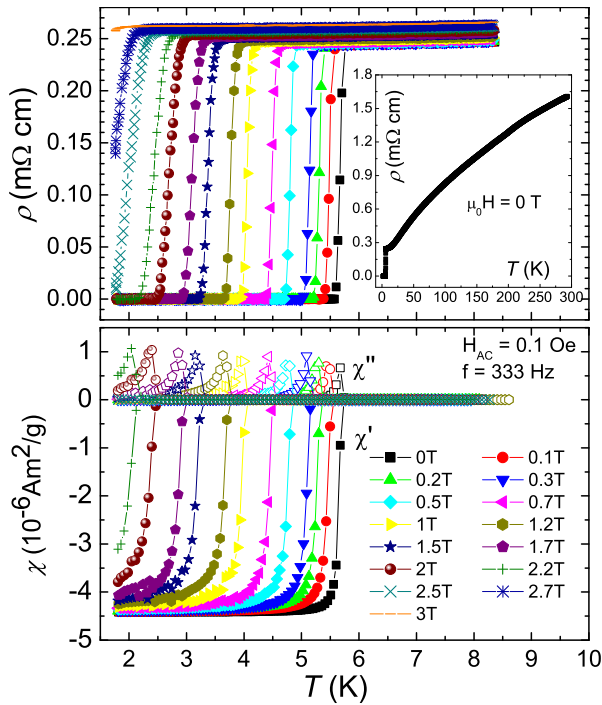}
\caption {(Color online) Temperature dependence of resistivity (top)
and magnetic susceptibility ($\chi ''$ and $\chi '$) (bottom) under
different dc magnetic fields. It is clear that the dc magnetic field
makes the transition shift parallel to low temperatures, manifesting
a field-induced pair-breaking effect. The inset in the top panel
shows the resistive transition in a wide temperature regime at
$\mu_0H$ = 0 T.} \label{fig2}
\end{figure}

As shown in Fig.1, the x-ray diffraction (XRD) pattern taken on one
sample with $T_c$ = 5.7 K shows a single phase with very small
amount of impurity which is comparable to that reported
previously\cite{Klimczuk}. After the first round of synthesizing,
the superconducting transition inspected by the ac susceptibility
occurs at about 5 K with a relatively wide transition. However,
after the second step, the transition moves to about 5.7 K with a
sharper transition width if we did not add extra Mg in the second
step. The Ta foil seems unreacted with the materials in both steps
of the fabrication if the Mg content is below 12 in the
stoichiometric formula.

In the top frame of Fig. 2 we show the temperature dependence of
resistivity under different magnetic fields. One can see that the
transition width determined from resistive measurements ($1\% - 99\%
\rho_n$ ) is only about 0.2 K. This is consistent with the rather
sharp magnetic transition as revealed by the ac susceptibility data
shown in the bottom view of Fig. 2. By applying a magnetic field the
transition shifts to lower temperatures quickly with a slope
$-d\mu_0H_{c2}(T)/dT|_{T_c} \approx$ 0.63 T / K for the present
sample with $T_c$ = 5.7 K. Using the Werthamer-Helfand-Hohenberg
relation\cite{WHH} $\mu_0H_{c2}=-0.69d \mu_0H_{c2}(T)/dT|_{T_c}
T_c$, we get the upper critical field $\mu_0H_{c2}$ = 2.48 T. The
value found here is comparable to that reported in the earlier
paper\cite{Klimczuk} ($T_c \approx $ 5 K). It is interesting to note
that the value of the slope $-d\mu_0H_{c2}(T)/dT$ correlates
strongly with $T_c$. For a sample with $T_c$ = 3.7 K, we found that
$-d \mu_0H_{c2}(T)/dT|_{T_c} \approx $ 0.2 T / K. This suggests that
there is a tunability of superconducting properties in this system
which will be addressed in Sec. IV.

\begin{figure}
\includegraphics[width=8cm]{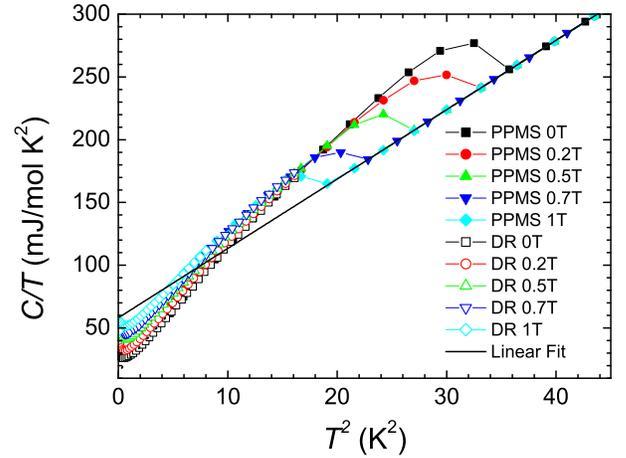}
\caption{(Color online) Raw data of specific heat plotted as $C/T$
vs $T^2$. All filled symbols represent the data taken with PPMS at
various magnetic fields. The open symbols show the data taken with
the DR. The thick solid line represents the normal state specific
heat which contains both the phonon and the electronic
contributions.} \label{fig3}
\end{figure}

\section{Specific heat and fit to the BCS model}

Shown in Fig. 3 are the raw data of the specific heat up to a
magnetic field of 1 T. The open symbols represent the data taken
with the DR, while all filled symbols show the data taken with the
PPMS. Both sets of data coincide very well for different fields.
With increasing the magnetic field the specific heat jump due to the
superconducting transition moves quickly to lower temperatures
leaving a background which is consistent with that above $T_c$ at
zero field. This provides a reliable way to extract the normal state
specific heat as shown by the thick solid line since the normal
state can be described by $C/T=\gamma_n +\beta T^2$, where the first
and the second term correspond to the normal state electronic and
phonon contribution, respectively. From the data it is found that
$\beta$ = 5.4 mJ/ mol K$^4$ and $\gamma_n$ = 57.7 mJ/mol K$^2$. By
extrapolating the data in the superconducting state at zero field
down to 0 K one finds, however, a residual value $\gamma_0 \approx$
13.8 mJ/ mol K$^2$ indicating a contribution either by a
nonsuperconducting fraction in volume of about 24\%, or by the DOS
induced by the impurity scattering for a nodal gap which will be
discussed later. The residual $\gamma_0$ may be induced by the
nonsuperconducting region, but it is still difficult to be regarded
as due to an impurity phase with completely different structure as
MgIrB since the XRD data shown in Fig. 1 is quite clean. We thus
suggest that the superconductivity depends sensitively on the
relative compositions among the three elements and some regions
without superconductivity have the chemical composition and even the
structure close to the superconducting phase. In any case, it is
safe to conclude that the normal state Sommerfeld constant for the
present sample is close to or slightly above 43.9 mJ/mol K$^2$.

\begin{figure}
\includegraphics[width=8cm]{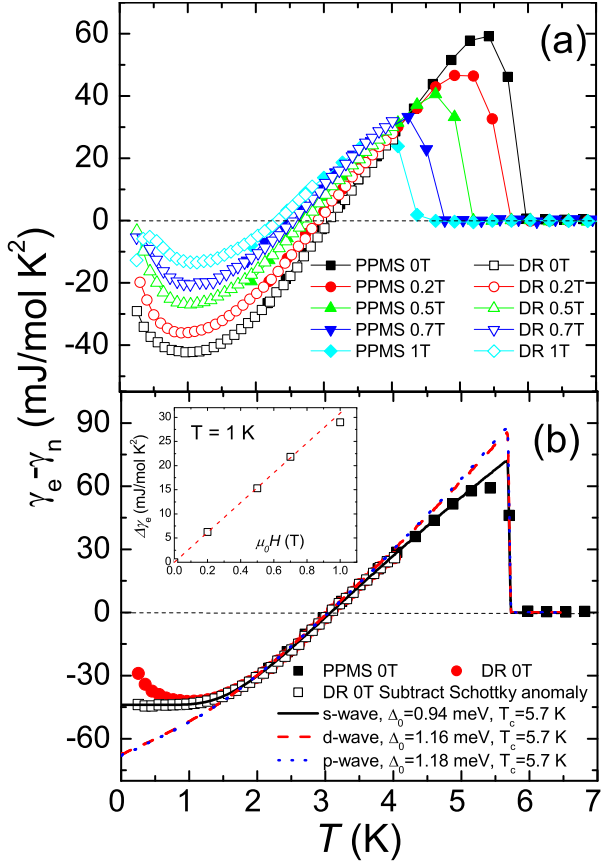}
\caption {(Color online) Temperature dependence of
$\gamma_e-\gamma_n$ for (a) magnetic fields up to 1 T and (b) at
zero field before (filled circles) and after removing (open and
filled squares) the Schotkky anomaly. The dark solid, red dashed,
and dotted green lines in (b) are the theoretical curves calculated
based on the BCS model (see text) with a gap of $s$ wave, $d$ wave,
and $p$ wave, respectively. } \label{fig4}
\end{figure}

Next we can have an estimation on the electron-phonon coupling
strength based on $\gamma_n$. In MgIrB, the electronic conduction is
dominated by the 5d band electrons of Ir atoms. The DOS at $E_F$
given by the local-density approximation band-structure
calculation\cite{Wiendlocha} is about $N(E_F)$ = 5.51/eV spin. By
assuming that there is an electron-phonon coupling constant
$\lambda_{e-ph}$ in the system, one has

\begin{equation}
\gamma_n=\frac{2\pi^2}{3}N(E_F)k_B^2(1+\lambda_{e-ph}).\label{eq:1}
\end{equation}

Using the band-structure value of $N(F_F)$, we have $\gamma_n$ =
25.98 $(1+\lambda_{e-ph})$(mJ/mol K$^2$). Taking the experimental
value $\gamma_n$ = 43.9 mJ/molK$^2$, we obtain $\lambda_{e-ph}
\approx$ 0.68, indicating a slightly moderate electron-phonon
coupling.

In the raw data shown in Fig. 3, one can see an upturn of
$\gamma=C/T$ in the very low-temperature region. This upturn is
known as the Schottky anomaly, induced by lifting the degeneracy of
the states of the paramagnetic spins. We tried a two level (S=1/2)
model to fit the low-temperature data but found a poor fitting
together with an extremely large Land\'{e} factor $g$ in the Zeeman
energy $g\mu_BH_{eff}$, where $\mu_B$ is the Bohr magneton,
$H_{eff}$ = $\sqrt{H^2+H_0^2}$ is the effective magnetic field which
evolves into $H_{eff}$ = $H_0$, the crystal field at zero external
field. In MgIrB the most possible paramagnetic centers may be from
Ir$^{4+}$ (S=5/2) or Ir$^{3+}$ (S=2). The system energy due to
Zeeman splitting in a magnetic field is\cite{Schottky}

\begin{equation}
E_{Sch}=\sum E_i\exp(-E_i/k_BT)/\sum \exp(-E_i/k_BT),\label{eq:3}
\end{equation}

where $E_i=M_Jg\mu_BH_{eff}$ and $M_J$ = $-$S, $-$S+1, ..., S$-$1,
S. The specific heat due to the Schottky effect is thus
$C_{Sch}=(n/k_B)dE_{Sch}/dT$, where $n$ represents the concentration
of the paramagnetic centers. For $S=5/2$ and $S=2$ the calculated
results are very close to each other, therefore we show only the fit
with $S=5/2$ corresponding to Ir$^{4+}$ (six levels). This method
allows us to deal with the data at zero and finite fields
simultaneously. It is known that the Schottky term should be zero at
$T$ = 0 K. In the superconducting state, the total specific heat can
be written as $C_{tot}=C_{nons}+C_e+C_{ph}+C_{Sch}$ with
$C_{nons}=\gamma_0T$ as the contribution of the nonsuperconducting
regions, $C_e$ is the electronic part. In the zero temperature limit
only the contribution of the nonsuperconducting part is left.
Applying a magnetic field gives rise to a finite value $\Delta
\gamma_e$ to $C_e=\gamma_eT$ due to the presence of vortices.
Practically, in order to fit the Schottky term, we first remove the
phonon contribution $C_{ph}=\beta T^3$, then vertically move the
experimental data downward with a magnitude $\gamma_0$ = 13.8 mJ/
mol K$^2$ and a field-induced vortex term $\Delta \gamma_e(H)$. In
Fig. 4(a) we present the specific heat coefficient
$\gamma_e-\gamma_n$ measured at the magnetic fields up to 1 T. An
upturn due to the Schottky effect is visible in the low temperature
region for all fields. In Fig. 4(b) we show the data at $\mu_0H$ = 0
T before [(red) filled circles] and after (dark open squares)
removing the Schottky anomaly calculated based on the model of six
levels with $\mu_0H_0$ = 0.25 T, n = 3.74 mJ/mol K and g = 2. Note
that we used the value 43.9 mJ/ mol K$^2$ as the normal state
Sommerfeld constant $\gamma_n$. One can see that the low-temperature
part after removing the Schottky anomaly is flattened out below
about 0.8 K when the field is zero. Furthermore, it can also be
justified by the requirement of entropy conservation. Since the
Schottky term gives only a very small contribution in the
high-temperature region (above 1.5 K here), if $\gamma_e$ had a
power law as required by a nodal gap, instead of a flat temperature
dependence for an $s$-wave gap, the entropy would be clearly not
conserved yielding a large negative entropy. This is of course
unreasonable. In Fig. 4(b) we present together the theoretical
curves for $\gamma_e - \gamma_n$ calculated using the weak coupling
BCS formula

\begin{eqnarray}
\gamma_\mathrm{e}=\frac{4N(0)}{k_BT^{3}}\int_{0}^{\hbar\omega_D}\int_0^{2\pi}\frac{e^{\zeta/k_BT}}{(1+e^{\zeta/k_BT})^{2}}(\varepsilon^{2}+\nonumber\\
\nonumber\\
\Delta^{2}(\theta,T)-\frac{T}{2}\frac{d\Delta^{2}(\theta,T)}{dT})\,d\theta\,d\varepsilon,
\end{eqnarray}

where $\zeta=\sqrt{\varepsilon^2+\Delta^2(T,\theta)}$. In obtaining
the theoretical fit we take the implicit relation $\Delta_0(T)$
derived from the weak coupling BCS theory for superconductors with
different pairing symmetries: $\Delta(T,\theta)=\Delta_0(T)$ for $s$
wave, $\Delta(T,\theta)=\Delta_0(T) cos2\theta$ for $d$ wave, and
$\Delta(T,\theta)=\Delta_0(T) cos\theta$ for $p$ wave, respectively.
The theoretical curve of $s$ wave fits the experimental data very
well leading to an isotropic gap value $\Delta_0$ = 0.94 meV and
$T_c$ = 5.7 K. The ratio $\Delta_0/k_BT_c$ = 1.91 obtained here is
quite close to the prediction for the weak coupling limit
($\Delta_0/k_BT_c$ = 1.76), indicating a slightly moderate coupling
strength. This is self-consistent with the conclusion derived from
the estimation on $\gamma_n$. In addition, the specific heat anomaly
at $T_c$ is $\Delta C_e/\gamma_nT|_{T_c} \approx $ 1.64 being close
to the theoretical value 1.43 predicted for the case of weak
coupling, again showing a slightly moderate coupling. The inset in
Fig. 4(b) shows a field induced part $\Delta \gamma_e$. In an
$s$-wave superconductor $\Delta \gamma_e$ is mainly contributed by
the vortex cores and a linear relation $\Delta \gamma_e \propto H
\gamma_n/H_{c2}(0)$ is anticipated\cite{Hussey} in the low field
region with $\Delta_0^2/E_F \ll T \ll T_c$. This linear relation is
well demonstrated by the data below 0.7 T, indicating another
evidence of $s$-wave pairing symmetry. This is in sharp contrast
with the results in cuprates where a $\Delta \gamma_e \propto
\sqrt{H}$ relation is expected due to the Doppler shift of the nodal
quasiparticle spectrum in a $d$-wave
superconductor\cite{Moler,WenHH}. The theoretical curves calculated
with $d$-wave and $p$-wave models cannot describe the data in both
the low-temperature and high-temperature
regimes\cite{Hussey,Volovik}.

\begin{figure}
\includegraphics[width=8cm]{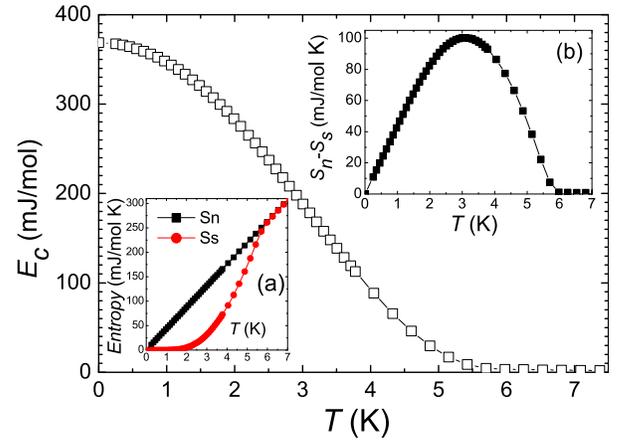}
\caption {(Color online) The main frame shows the superconducting
condensation energy calculated from the specific heat. Inset (a)
shows the entropy in the normal and superconducting state. Plotted
in inset (b) is the difference of the entropy between the normal and
superconducting state. } \label{fig5}
\end{figure}

In the following we try to estimate the superconducting condensation
energy $E_{c}$ for the sample with $T_c$ = 5.7 K. In calculating
$E_{c}$ we get the entropy difference between the normal state and
the superconducting state by $S_n-S_s=\int_0^T(\gamma_n-\gamma_e)d
T'$, then $E_{c}$ is calculated through $E_{c}=\int_T^{6
K}(S_n-S_s)dT'$. The data of $S_n$ and $S_s$ as well as the
difference between them are shown in insets (a) and (b) of Fig. 5,
respectively. The main frame of Fig. 5 shows the temperature
dependence of the condensation energy $E_c$ which is about 369
mJ/mol at $T$ = 0 K. This value can actually be assessed by the
following equation

\begin{equation}
E_{c}=\alpha N(E_F)\Delta_0^2/2=\alpha
\frac{3}{4\pi^2}\frac{1}{k_B^2}\gamma_n\Delta_0^2.\label{eq:4}
\end{equation}

For a BCS $s$-wave superconductor, $\alpha$ = 1, taking $\gamma_n$ =
43.9 mJ/mol K$^2$ and $\Delta_0$ = 0.94 meV, we found a value of
$E_{c} \approx $ 367 mJ/mol which is remarkably close to the
experimental value 369 mJ/mol. This also validates the values of
$\Delta_0$ and $\gamma_n$ determined in our experiment.

Now we discuss the possibility of impurity scattering which may
suppress the spin-triplet component of the superconductivity.
Theoretically it is known that the nodal gap structure is very
sensitive to the impurities. If the spin-singlet and triplet
components are mixed, the latter might be suppressed by the impurity
scattering and the system would behave like a BCS superconductor.
Although we cannot exclude this possibility at this moment, it is
safe to conclude that the spin-triplet component (if it exists)
should be a very small part of the total condensate. This can be
justified by the following two arguments. First, for the sample with
$T_c$ = 5.7 K, we found that $\gamma_0$/$\gamma_n(T > T_c)$ = 24\%.
Assuming that the total $\gamma_0$ were induced by the impurity
scattering of this spin-triplet component, the ratio of the DOS
corresponding to the normal state of this part would be less than
24\%. Actually one would not expect that the total value of
$\gamma_0$ here could be ascribed completely to the suppressed
spin-triplet component since the XRD data does show the existence of
some secondary (nonsuperconducting) phase which will certainly
contribute a certain value of DOS. Second the self-consistency among
the derived values of the normal state Sommerfeld constant, the
appropriate ratio of $\Delta_0/k_BT_c$ and the electron-phonon
coupling strength (see next section) together with the nice fit of
the specific heat data to the BCS model all can push the
spin-triplet component to a very low limit. Nevertheless, as a
future work it is worthwhile to have an inspection on this issue on
a sample with complete purity.

\section{Electron-phonon coupling strength and the tunability of superconductivity in MgIrB}
\begin{figure}
\includegraphics[width=8cm]{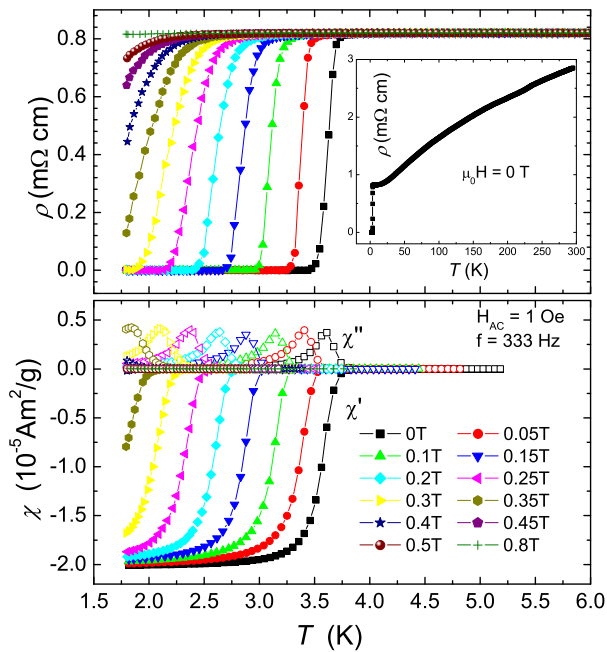}
\caption {(Color online) Temperature dependence of resistivity (top)
and magnetic susceptibility ($\chi''$ and $\chi'$) (bottom) under
different dc magnetic fields for the sample with $T_c$ = 3.7 K. It
is clear that the dc magnetic field makes the transition shift
parallel to low temperatures with a much faster speed compared with
the sample with $T_c$ = 5.7 K. } \label{fig6}
\end{figure}

\begin{figure}
\includegraphics[width=8cm]{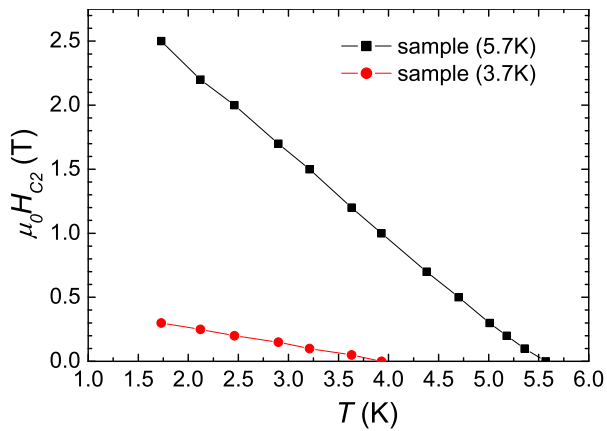}
\caption {(Color online) The upper critical field for both samples
with different transition temperatures.  } \label{fig7}
\end{figure}
Now we get to the electron-phonon coupling in MgIrB. From the normal
state value we have derived the phonon contribution $C_{ph}=\beta
T^3$ with $\beta \approx$ 5.4 mJ/mol K$^4$. Using the relation
$\Theta_D$ = $(12\pi^4k_BN_AZ/5\beta)^{1/3}$, where $N_A$ = 6.02
$\times 10^{23}$ is the Avogadro constant, Z = 45 is the number of
atoms in one unit cell, we get the Debye temperature $\Theta_D
$(MgIrB) = 253 K for the sample with $T_c$ = 5.7 K. If the
conduction electrons are weakly coupled with the phonon, one can use
the McMillan equation to evaluate the electron-phonon coupling
strength $\lambda_{e-ph}$ via\cite{McMillan}

\begin{equation}
T_c=\frac{\Theta_D}{1.45}\exp[-\frac{1.04(1+\lambda_{e-ph})}{\lambda_{e-ph}-\mu^*(1+0.62\lambda_{e-ph})}],
\end{equation}

where $\mu^*$ is the Coulomb pseudopotential taking 0.11. Using
$\Theta_D$(MgIrB) = 253 K and $T_c$ = 5.7 K, we obtain
$\lambda_{e-ph}$ = 0.66, indicating a slightly moderate coupling
strength. The value determined here is quite close to
$\lambda_{e-ph}$ = 0.68 which was determined previously based on the
comparison between the normal state Sommerfeld coefficient
$\gamma_n$ and the theoretically estimated DOS at $E_F$. All these
detailed and self-consistent analysis indicate an electron-phonon
coupling mechanism for superconductivity in MgIrB.

At this moment, it is still unclear that the conduction electrons
are strongly coupled to which phonon branch leading to the
superconductivity. During the preparation of samples, it was found
that the superconducting transition temperature is changeable in a
wide region (from 2 K to 5.7 K) in MgIrB with a rather sharp
transition width. This is actually a rare case in the alloy
superconductors in which the off-stoichiometric elements normally
play as the scatterers and break the Cooper pairs and finally make
the transition broad. In MgIrB a slight off-stoichiometry may not
change the structure, but promote the mutual-substitution leading to
a different electron-phonon coupling strength. In Fig. 6 we show the
temperature dependence of the resistivity and the ac susceptibility
of another sample ($T_c$ = 3.7 K) which was made also in two steps
but with 20\% more Mg added in the second step. One can see that the
transition width at zero field is still quite narrow and the x-ray
diffraction pattern taken on this sample is reasonably clean (not
shown here). An estimation on the electron-phonon coupling strength
$\lambda_{e-ph}$ based on the McMillan equation on this sample tells
that $\lambda_{e-ph} \approx $ 0.55. In Fig. 7 we present the upper
critical field $H_{c2}(T)$ taken with the criterion 10\%$\rho_n$ for
the two selected samples with $T_c$ = 5.7 K and $T_c$ = 3.7 K. One
can see that the slope $-d\mu_0H_{c2}/dT|_{T_c}$ is very different
between these two samples. Actually in the previous work of the
Princeton group\cite{Klimczuk}, $-d\mu_0H_{c2}/dT|_{T_c}$ can be as
high as 1 T/K. In a dirty type-II superconductor, it was
predicted\cite{JaffePRB1989} that $-d \mu_0H_{c2}/dT|_{T_c}\propto
\rho_n\gamma_n \eta$ with $\eta$ in connection with the
electron-phonon coupling strength. It is thus reasonable to ascribe
the variation of $d\mu_0H_{c2}/dT|_{T_c}$ and $T_c$ to different
electron-phonon coupling strengths and different density of states
at $E_F$. All these quantities may be optimized in future work and
hopefully will lead to a higher superconducting transition
temperature. The basic parameters and properties derived in this
work provide a playground for the future study in this interesting
system.

\section{Concluding remarks}
In summary, analysis on the low-temperature data in MgIrB finds an
$s$-wave pairing symmetry with a gap in the weak coupling to
slightly moderate coupling regime and the superconductivity is
induced by the electron-phonon coupling. The spin-triplet component,
if it exists, should remain as a small part of the total condensate.
Tuning the superconducting transition temperature seems possible
through changing the electron-phonon coupling strength and the
density of states at the Fermi level by varying the relative
compositions among the three elements and probably also by mutual
substitution.

\begin{acknowledgments}
The authors acknowledge the fruitful discussions with Tao Xiang and
Junren Shi at IOP, CAS, and Guoqing Zheng at Okayama University,
Japan. This work is supported by the National Science Foundation of
China, the Ministry of Science and Technology of China (973 project,
Contracts Nos. 2006CB601000 and 2006CB921802), and Chinese Academy
of Sciences (project ITSNEM).
\end{acknowledgments}

\end{document}